# Spin-affected reflexive and stretching separation of off-center droplet collision

Chengming He[1], Lianjie Yue[1], and Peng Zhang[2,*]

[1]Institute of Mechanics, Chinese Academy of Sciences, No.15 Beisihuanxi Road, Beijing, China, 100190
[2]Department of Mechanical Engineering, The Hong Kong Polytechnic University, Hum Hong, Kowloon, Hong Kong, 999077

Recent studies have demonstrated the significant roles of droplet self-spin motion in affecting the head-on collision of binary droplets. In this paper, we present a computational study by using the Volume-of-Fluid (VOF) method to investigate the spin-affected droplet separation of off-center collisions, which are more probable in reality and phenomenologically richer than head-on collisions. Different separation modes are identified through a parametric study with varying spinning speed and impact parameter. A prominent finding is that increasing the droplet spinning speed tends to suppress the reflexive separation and to promote the stretching separation. Physically, the reflexive separation is suppressed because the increased rotational energy reduces the excessive reflexive kinetic energy within the droplet, which is the cause for the droplet reflexive separation. The stretching separation is promoted because the increased droplet angular momentum enhances the local stretching flow within the droplet, which tends to separate the droplet. The roles of orbital angular momentum and spin angular momentum in affecting the droplet separation are further substantiated by studying the collision between two spinning droplets with either the same or opposite chirality. In addition, a theoretical model based on conservation laws is proposed to qualitatively describe the boundaries of coalescence-separation transition influenced by droplet self-spin motion.

---

* Corresponding author
E-mail address: pengzhang.zhang@polyu.edu.hk (P. Zhang)

## I. INTRODUCTION

Collision between two droplets in a gaseous environment is of relevance to many natural and industrial processes, and it has been extensively studied[1-14] and reviewed[15, 16] in the literature. For the simplest situation involving two identical droplets in atmosphere, previous experimental studies[1, 3, 5, 10, 12-14] identified and interpreted some distinct collision outcomes, such as (I) coalescence, (II) bouncing, (III) coalescence, (IV) reflexive separation, and (V) stretching separation. These outcomes are schematized by a well-known collision nomogram in the two-dimensional $We - B$ parameter space, as shown in Fig. 1. The collision Weber number, $We = \rho_l D_l U^2/\sigma$ ($\rho_l$ is the density of the liquid, $D_l$ the droplet diameter, $U$ the relative velocity, and $\sigma$ the surface tension), measures the relative importance of the droplet impact energy compared to the surface energy and is limited to be of $O(10^2)$, beyond which more complex collision outcomes emerge[9, 17]. The impact parameter, $B = \chi/D_l$ ($\chi$ is the projection of mass center connection line in the direction perpendicular to $U$), measures the deviation of the trajectory of droplets from that of head-on collision, with $B = 0$ denoting head-on collision and $B = 1$ grazing collision. In this nomogram, the droplet Ohnesorge number, $Oh = \mu_l/\sqrt{\rho_l D_l \sigma}$ (where $\mu_l$ is the dynamic viscosity of the liquid), which measures the relative importance of the liquid viscous stress compared to the capillary pressure, is often not fixed but varies within a small range of values. To obtain a complete parametric description, the Ohnesorge number must be treated as an independent controlling parameter[8, 12, 14]. For situations more realistic than that of identical droplets in atmosphere, additional physical parameters should be considered, and a higher-dimensional parameter space may be needed for establishing a comprehensive collision nomogram. For example, the size ratio, $\Delta$, measures the influence of droplet size disparity[3, 10, 11]; the gas-liquid density ratio (or equivalent dimensionless parameter) to measure the gas pressure effect[5, 18, 19]; the surface tension ratio to measure the Marangoni effect for collision of droplets of unlike liquids[20-23].

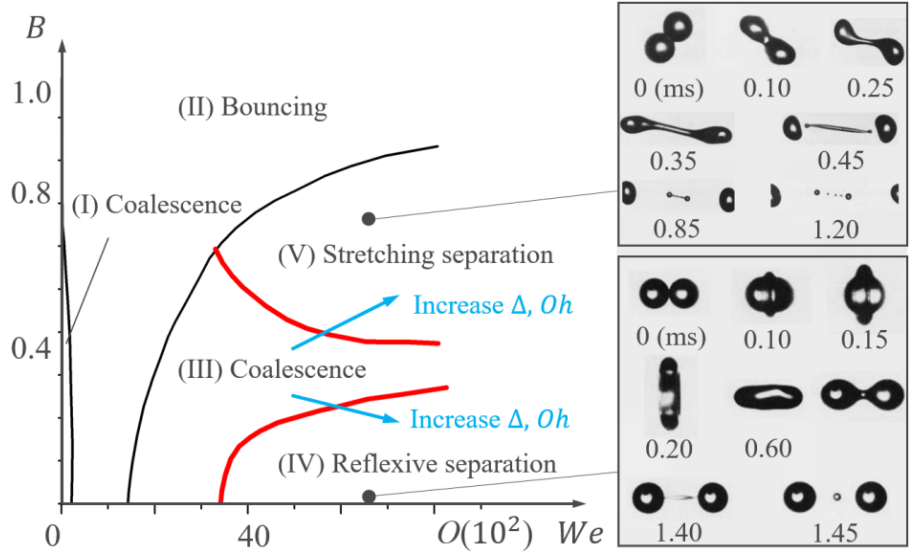

FIG. 1. Schematic of droplet collision regimes in a $We - B$ parameter space and experimental images for reflexive and stretching separation[4] for collision of identical droplets in atmosphere.

Among various droplet outcomes, droplet coalescence and separation are often of great interest because they may significantly change the droplet size distribution in dispersed liquid-gas flows. Given other collision parameters being fixed, two colliding droplets with a small impact kinetic energy (i.e. at small $We$) tend to merge into a larger droplet. Although the colliding droplets may bounce back at intermediate $We$, they tend to separate into two large bulks of liquid with a certain number of satellite droplets if the impact kinetic energy is substantially larger than the surface energy (i.e. at sufficiently large $We$)[3-5]. As increasing $Oh$ and $\Delta$, the regime boundaries between coalescence and separation tend to move towards a larger $We$, because the augmented viscous dissipation of the droplet internal flow stabilizes the merged droplet[11, 12, 24]. The droplet separation and satellite droplet formation have been found to be important physics components in modelling various spray processes[25-27].

The regimes of coalescence and separation (including reflexive separation and stretching separation) are shown in Fig. 1. In a certain range of $We$, droplet collision outcomes show a non-monotonic variation from reflexive separation to coalescence and to stretching separation as increasing $B$ [3-6, 10, 11]. For collisions at small $B$[4, 5], the reflexive separation (regime IV) occurs when the kinetic energy of droplet collision is substantially larger than the surface energy of droplets. The droplets may still possess a significant amount of translational kinetic energy upon their coalescence. If the viscous internal flow within the coalesced droplet does not rapidly dissipate the excessive kinetic energy, the coalesced droplet cannot be held by its surface tension and will be elongated in the direction of droplet collision to result in a separation. For collisions at larger $B$, the colliding droplets tend to form a long ligament by the stretching flow motion, and the stretching separation (regime V) occurs when the kinetic energy of liquid ligament is larger than the sum of surface energy and stretching-flow-induced viscous dissipation[3, 4]. Furthermore, for collisions at intermediate $B$ and $We$ of $O(10^2)$ or higher, a new regime named "rotational separation" were identified[28], which was interpreted by the balance between rotational energy and surface energy[1, 4].

In realistic situations droplet collision is more complicated than that in the cases discussed above due to the presence of droplet self-spin motion. The previous experimental[3, 4] and numerical[29-31] studies have demonstrated that two initially non-spinning droplets undergoing off-center collisions (leading to either bouncing or coalescence) can generate significant spinning motion of droplets. This is because the initial orbital angular momentum of the two droplet system[30, 31] can be partially converted into the spin angular momentum of each droplet. It is very likely that every droplet is a real spray system would possess a spin motion of certain speed if droplet interaction is present in the system.

The previous understanding of droplet collision would be modulated by considering the presence of droplet spin. For the head-on bouncing between spinning droplets[32], the spinning droplet can induce significant non-axisymmetric droplet deformation because of the conversion of the spin angular momentum into the orbital angular momentum. The interchange between orbital and spin angular momentums during the collision process is of significance because it can influence the post-collision velocities of bouncing droplets. For head-on coalescence between a spinning droplet and a non-spinning droplet of equal size[32], the spinning motion can promote the mass interminglement of droplets because the locally nonuniform mass exchange occurs at the early collision stage by non-axisymmetric flow and is further stretched along the filament at later collision stages. Apart from the study of spinning effects on droplet bouncing and coalescence, the spin-affected droplet separation and subsequent satellite droplet formation are highly probable in practical dense sprays, but relevant studies have not been reported in the literature.

In this paper, we present a computational study to investigate the spinning effects on the droplet separation. The presentation of the study is organized as follows. The numerical methodology and specifications are described in Sec. II. The separation modes and mechanisms affected by impact parameter and droplet spinning speed are presented in Sec. III, followed by the influences of chirality of droplet self-spin motion on reflexive and stretching separations in Sec. IV. Finally, a theoretical model for reflexive and stretching separation is proposed to account for the boundary transition influenced by the chirality of droplet spin motion in Sec. V.

## II. COMPUTATIONAL METHODOLOGY AND SPECIFICATIONS

### A. Methodology and validations

The three-dimensional (3D) continuity and incompressible Navier-Stokes equations,

$$\nabla \cdot \boldsymbol{u} = 0 \tag{1}$$

$$\rho(\partial \boldsymbol{u}/\partial t + \boldsymbol{u} \cdot \nabla \boldsymbol{u}) = -\nabla p + \nabla \cdot (2\mu \boldsymbol{D}) + \sigma \kappa \boldsymbol{n} \delta_s \tag{2}$$

are solved by using the classic fractional-step projection method, where $\boldsymbol{u}$ is the velocity vector, $\rho$ the density, $p$ the pressure, $\mu$ the dynamic viscosity, and $\boldsymbol{D}$ the strain rate tensor defined as $D_{ij} = (\partial_j u_i + \partial_i u_j)/2$. In the surface tension term $\sigma \kappa \boldsymbol{n} \delta_s$, $\delta_s$ is a Dirac delta function, $\sigma$ the surface tension coefficient, $\kappa$ the local curvature, and the unit vector $\boldsymbol{n}$ normal to the local interface.

To solve both the gas and liquid phases, the density and viscosity are constructed by the volume fraction as $\rho = c\rho_l + (1-c)\rho_g$ and $\mu = c\mu_l + (1-c)\mu_g$, in which the subscripts $l$ and $g$ denote the liquid and gas phases, respectively. The volume fraction $c$ satisfies the advection equation

$$\partial c/\partial t + \nabla \cdot (c\boldsymbol{u}) = 0 \tag{3}$$

with $c = 1$ for the liquid phase, $c = 0$ for the gas phase, and $0 < c < 1$ for the gas-liquid interface. The present study adopts the Volume-of-Fluid (VOF) method, which has been implemented in the open source code, Gerris[33, 34], featuring the 3D octree adaptive mesh refinement, the geometrical VOF interface reconstruction, and continuum surface force with height function curvature estimation. Gerris has been demonstrated to be competent for high-fidelity simulation of a wide range of multiphase flow problems[29-32, 35-40].

A major challenge of VOF simulation of droplet collision lies in the absence of subgrid models describing the rarified gas effects and the Van der Waals force[41] within the gas film, thereby prohibiting the physically realistic prediction of droplet coalescence and separation. A coarse mesh would result in a "premature" coalescence of droplets without prominent droplet deformation. Thus, the successful simulation of droplet coalescence and subsequent collision dynamics in previous studies[29, 35, 38] were obtained by choosing an appropriate mesh resolution in the vicinity of the interface. However, it is noted that the resolution of gas film drainage is not a significant issue for the simulation of droplet separation at large $We$. This is because the drainage of gas film occurs at a much short time scale compared with that of the entire process of the coalescence and separation of droplets. In addition, the energy budget during the gas film drainage accounts for a very small portion of the total energy budget. As a result, the computational uncertainties of droplet coalescence caused by a relatively coarse mesh for gas film will not result in significant difference to the "long-time" dynamics of droplet deformation and separation. Chen *et al.*'s VOF simulation[29, 42] shows that a maximum interface mesh refinement of level 11 is required to produce the physically correct droplet bouncing at $We = 8.6$[29], but a mesh refinement of level 8 can produce satisfactory results of droplet separation at $We = 61.4$ [42]. Similarly, the level-set method[43] and lattice Boltzmann method (LBM)[21, 44] have also been used to simulate droplet separation with a reasonably refined mesh and without any subgrid models for rarified gas effects and the Van der Waals force.

To justify the computational approach discussed above, the experimental validation and grid independence analysis are shown in Fig. 2 for the case at $We = 61.4$, $B = 0.06$, and $Oh = 0.028$ of Qian and Law[5]. To improve computational efficiency, the entire computational domain is divided into three physical zones, namely the gas, the droplet, and the interface zones, and different mesh refinement level ($N_g, N_l, N_i$) is used in these zones. The simulation results by Chen *et al.*[42] with a mesh refinement level of (4,7,8) are shown in Fig. 2(b) for comparison. The present results with the same mesh refinement level are shown in Fig. 2(c). The presented simulation results are those having the mostly agreed droplet deformation with the experimental images, whereas the time discrepancies between them could serve as an indicator for simulation errors. It is seen that the experimental and simulation times display slight discrepancies throughout the entire collision process, and the time errors are less than 8%. A typical simulation run with the mesh refinement level (4, 7, 8) results in $3.73 \times 10^6$ grid points in the entire droplet, taking about 200 hours of real time to run the simulation up to T = 2.0 on two Intel Xeon(R) Gold-6150 processor with 72 cores (36 cores for each processor).

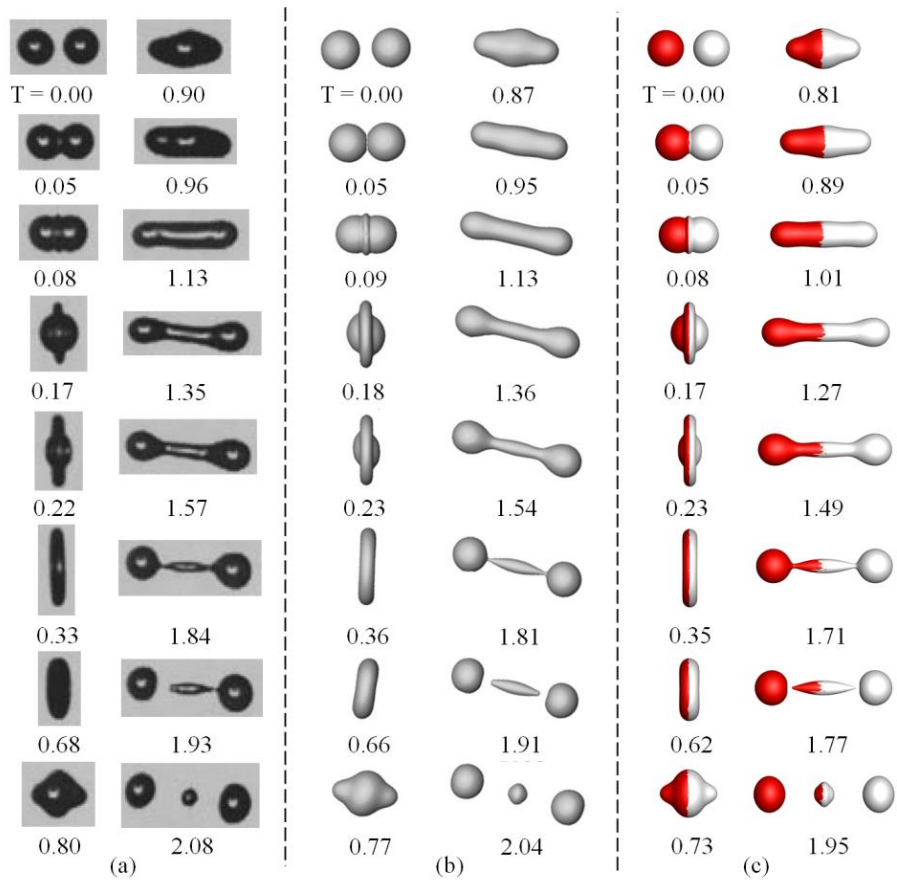

FIG. 2. Comparison of (a) experimental images from Qian and Law[5], (b) simulation results from Chen et al.[42], and (c) the present results, for the nearly head-on separation of identical droplets

at $We = 61.4$, $Oh = 0.028$, and $B = 0.06$. The dimensionless time $T = t/t_{osc}$ and $t_{osc} = (\rho_l R_l^3/\sigma)^{1/2} = 1.06$ ms.

**B. Problem description and numerical specifications**

The 3D computational domain of droplet collision is illustrated from different perspectives, as show in Fig. 3. Two droplets of diameter $D$ are specified to collide along the *x*-direction with a relative translational velocity, $U$, and therefore they have zero relatively velocities in the *y*- and *z*-directions. Without losing generality, the translational velocity component for droplet $O_1$ and $O_2$ are set as $-U/2$ and $U/2$ along the *x*-direction, respectively, so that the linear momentum of the entire mass-center system remains zero. For off-center collisions, the deviation of the mass centers from the head-on collision is qualified by $\chi$, which is defined as the projection of the connection line $O_1O_2$ (hereinafter referred to $\overline{O_1O_2}$) along the *z*-direction. The spin axis $l_{O_1}$ can be specified by a polar angle $\theta$ with respect to the *z*-axis and an azimuthal angle $\varphi$ to the *x*-axis. In our previous study[32], we deliberately fixed the polar angle $\theta = \pi/2$ and varied the azimuthal angle $\varphi$ in the range of $0 < \varphi < \pi/2$, because the situations with varying $\theta$ are equivalent for the axisymmetric head-on droplet collision. As a result, the initial spin angular velocity can be expressed as $\boldsymbol{\omega}_0 = (-\omega_0 cos\varphi, -\omega_0 sin\varphi, 0)$, and the spinning velocity components of droplet $O_1$ is given by $H(\phi - 1)\boldsymbol{\omega}_0 \times (\boldsymbol{r} - \boldsymbol{R}_{O_1})$, where $\phi$ is the color function with $\phi = 1$ in the spinning droplet $O_1$ and otherwise $\phi = 0$, and the Heaviside step function $H$ ensures the assignment of spin to droplet $O_1$ only. The domain is $6D$ in length and $4D$ in both width and height; all the boundaries are specified with the free outflow boundary conditions.

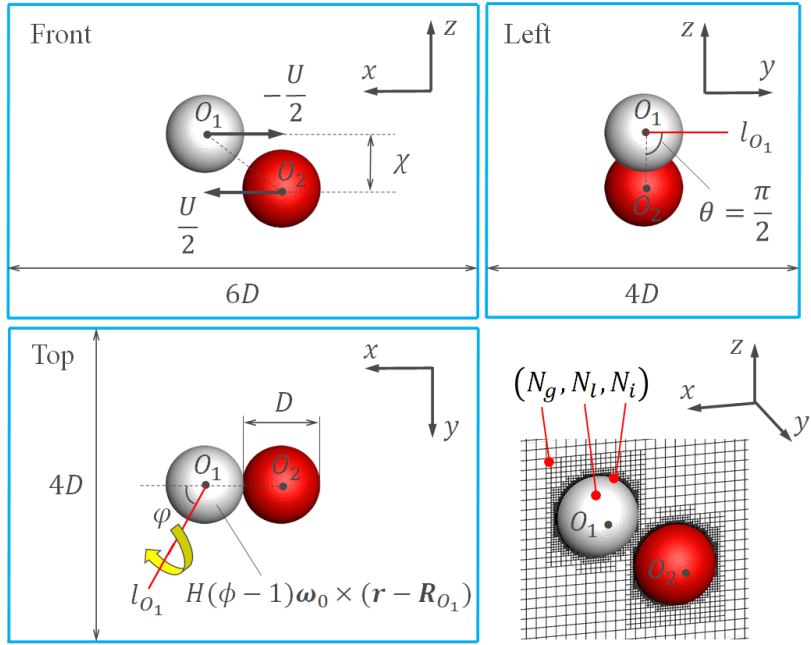

FIG. 3. Three-dimensional computational setup for an off-center collision between a spinning droplet $O_1$ and a non-spinning droplet $O_2$.

To simplify the parametric study in the present work, the polar angle $\theta$ is fixed at $\pi/2$ and the azimuthal angle $\varphi$ is fixed at $\pi/2$. This is because the non-axisymmetric flow characteristics induced by the droplet spin is the most prominent at $\varphi = \pi/2$, owing to the strongest interaction between the spinning motion and the translational motion[32]. This simplification is justified by examining the simulation results shown in Figure 4. The deformation of the head-on collision between two non-spinning droplets is axisymmetric by *x*-axis and "mirror symmetry" with respect to the (*y-z*) plane, as seen in Fig. 4(a). For collisions between a spinning droplet and a non-spinning droplet, as shown in Fig. 4(b-d) with varying $\varphi$ from 0 to $\pi/2$, the droplet deformation is deviated from the head-on collision owing to the conversion of spin angular momentum into orbital angular momentum. The droplet separation followed by the asymmetric pinch-off of the ligament is delayed when compared to that of non-spinning droplet collision owing to the enhanced viscous dissipation induced by the droplet's translational impacting motion and droplet spinning motion. The asymmetry from the z-plane reaches its maximum at $\varphi = \pi/2$, where there is the strongest interaction between the spinning motion and the translational motion.

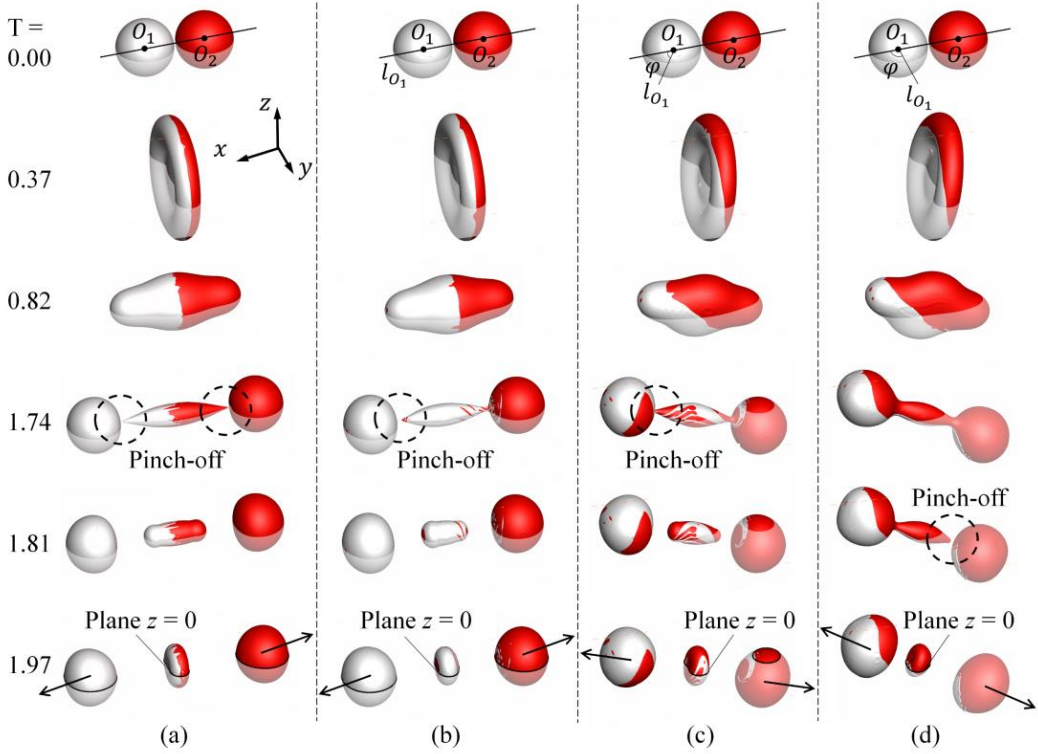

FIG. 4. Evolution of droplet deformation upon a head-on collision of (a) two non-spinning droplets and of a spinning droplet $O_1$ and a non-spinning droplet $O_2$ with varying azimuthal angle of droplet $O_1$ (b) $\varphi = 0$, (c) $\varphi = \pi/4$, and (d) $\varphi = \pi/2$, at $We = 61.4$, $Oh = 0.028$, and $\omega_0 = 3$.

In addition, to further simplify the problem but not to lose generality, this study restricts its scope to the collision between two equal-sized droplets ($\Delta = 1.0$) so as to avoid unnecessary complexity of geometrical asymmetry and size disparity. Furthermore, we recognized that the post-collision velocity vector of the separated droplets may be affected by the spinning axis ($\theta$ and $\varphi$), and the parametric study of spinning axis orientation will be considered in our future studies. Consequently, the present numerical study focuses on the controlling parameters in the range of $We = 40\sim85$, $B = 0\sim1.0$, $\omega_0 = 0\sim6$, and $Oh = 0.028$.

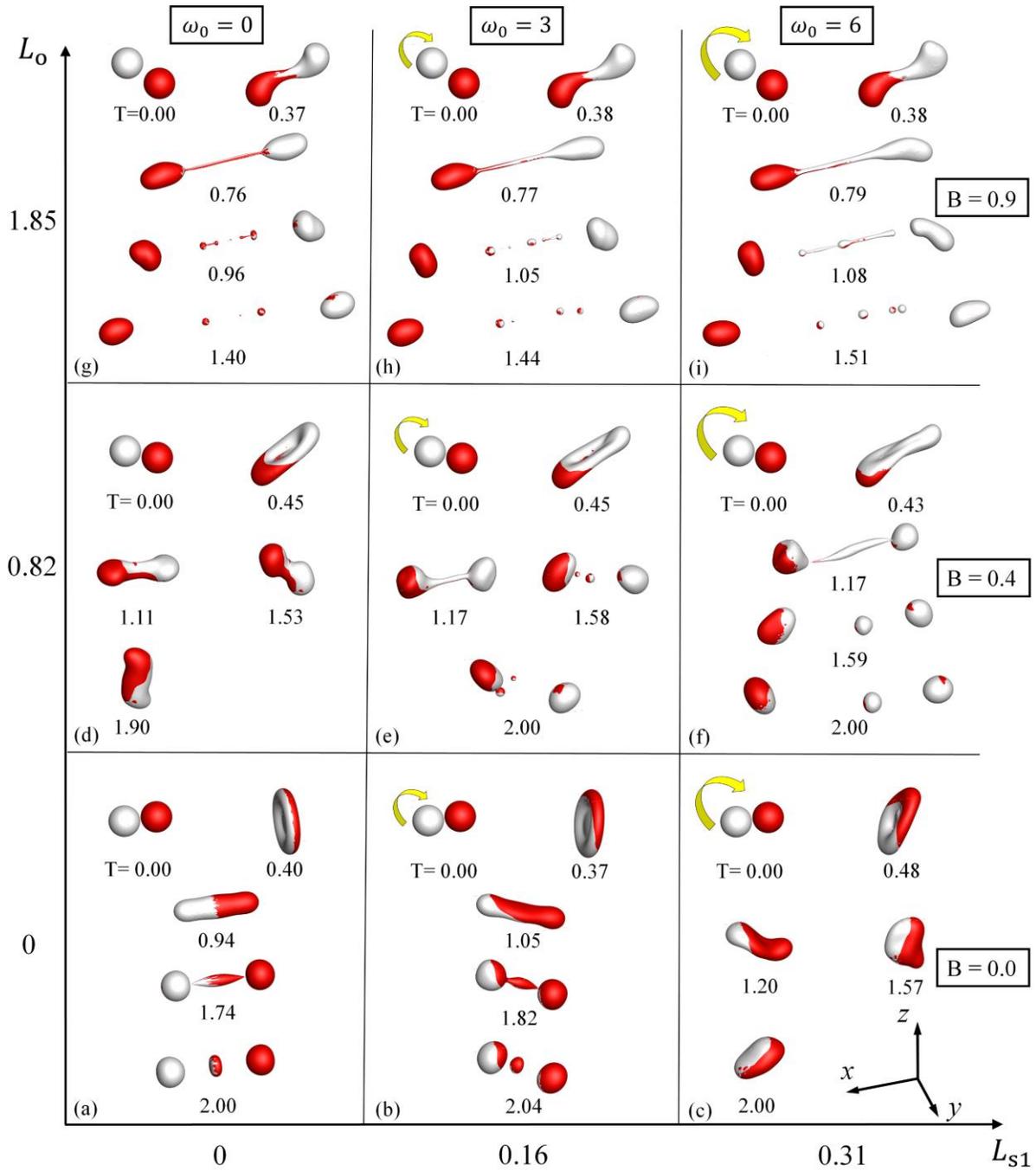

FIG. 5. Collision between a spinning droplet and a non-spinning droplet with varying spinning speed $\omega_0$ and impact parameter $B$ at $We = 61.4$ and $Oh = 0.028$.

## III. EFFECTS OF DROPLET SPIN SPEED AND IMPACT PARAMETER

### A. Separation modes affected by $\omega_0$ at different $B$

Figure 5 shows the droplet separation affected by different droplet spinning speed $\omega_0$ and impact parameter $B$. To facilitate the physical understanding from the perspective of angular momentum conversion, the consequence of varying $\omega_0$ and $B$ is also characterized as the change of spin angular momentum $\boldsymbol{L}_{s1}$ of droplet $O_1$ with respect to the spinning axis across its mass center and the orbital angular momentum $\boldsymbol{L}_o$ with respect to the *y*-axis for two droplets system, respectively.

The time-dependent angular momentum $\boldsymbol{L}_{s1}$ and $\boldsymbol{L}_o$ are calculated by

$$\boldsymbol{L}_{s1}(t) = \int_V \rho_l H(\phi - 1)(\boldsymbol{r} - \boldsymbol{R}_{O_1}) \times \boldsymbol{v} \, dV' \tag{4}$$

and

$$\boldsymbol{L}_o(t) = \int_V \left[\rho_l H(\phi - 1)\boldsymbol{R}_{O_1} + \rho_l H(c - \phi - 1)\boldsymbol{R}_{O_2}\right] \times \boldsymbol{v} \, dV' \tag{5}$$

where $H$ is the Heaviside step function, $\phi$ the color function used in VOF simulation, $c$ the volume fraction, $\boldsymbol{r}$ and $\boldsymbol{v}$ the position vector and velocity vector, $V$ the integral volume of liquid and gas phases. The time-dependent $\boldsymbol{R}_{O_1}$ and $\boldsymbol{R}_{O_2}$ are the position vectors of the mass centers $O_1$ and $O_2$ for two droplets and defined by

$$\boldsymbol{R}_{O_1}(t) = \int_V \rho_l H(\phi - 1)\boldsymbol{r} \, dV'/M_1 \tag{6a}$$

$$\boldsymbol{R}_{O_2}(t) = \int_V \rho_l H(c - \phi - 1) \boldsymbol{r}\, dV'/M_2 \tag{6b}$$

Although $\boldsymbol{L}_{s1}$ and $\boldsymbol{L}_o$ change with time during droplet collision, the conserved total angular momentum $\boldsymbol{L}_t = \boldsymbol{L}_o + \boldsymbol{L}_{s1}$ poses an additional restriction on the motion of droplets. Furthermore, owing to the polar angle $\theta$ and azimuthal angle $\varphi$ are fixed at $\pi/2$ in the present study, the angular momentum $\boldsymbol{L}_o$ and $\boldsymbol{L}_{s1}$ (and hence $\boldsymbol{L}_t$) have a nonzero component only in the $y$-direction. The $y$-components, $L_{s1}$ and $L_o$, at the initial time are given by

$$L_{s1}(t=0) = \frac{2}{5}\left(\frac{4}{3}\pi R^3 \rho_l\right) R^2 \omega_0 \tag{7a}$$

$$L_o(t=0) = \frac{1}{2}\left(\frac{4}{3}\pi R^3 \rho_l\right)(2R)UB \tag{7b}$$

Three cases of head-on collisions at $B = 0.0$ are shown in Fig. 5 (a-c), where the non-spinning case with $\omega_0 = 0$ is also shown as a benchmark case for comparison. It is seen that the benchmark case clearly shows the typical physical phenomena of reflexive separation. Specifically, the colliding droplets undergo the following stages: (I) the initial coalescence, (II) the radial deformation to form a pancake shape, (III) the retraction motion under surface tension towards the axis (the $x$-direction), (IV) the stretching motion along the axis owing to the droplet inertia, (V) the formation of a liquid ligament being attenuated at the ends, (VI) the final pinch-off of the ligament simultaneously at the two sufficiently attenuated ends, (VII) the contraction of pinched-off ligament under surface tension to form a satellite droplet. For the spinning case with $\omega_0 = 3$, the colliding droplets still undergo the same stages as described above but have apparent non-axisymmetric appearances. Particularly, the unsymmetric liquid ligament pinches off first from the attenuated end connected to the non-spinning droplet. Such a non-axisymmetric droplet collision becomes so significant for the case with $\omega_0 = 6$ that the droplet separation is completely suppressed.

For the collisions at intermediate $B = 0.4$, which are shown in Fig. 5(d-f), the non-spinning case shows no separation but permanent coalescence, but the stretching separation is manifested for the spinning cases. It is seen that a longer liquid ligament is formed as increasing $\omega_0$. For the collisions at larger $B = 0.9$, as shown in Fig. 5 (g-i), the formed ligament is elongated as increasing $\omega_0$, and it can break up into multiple satellite droplets of different sizes.

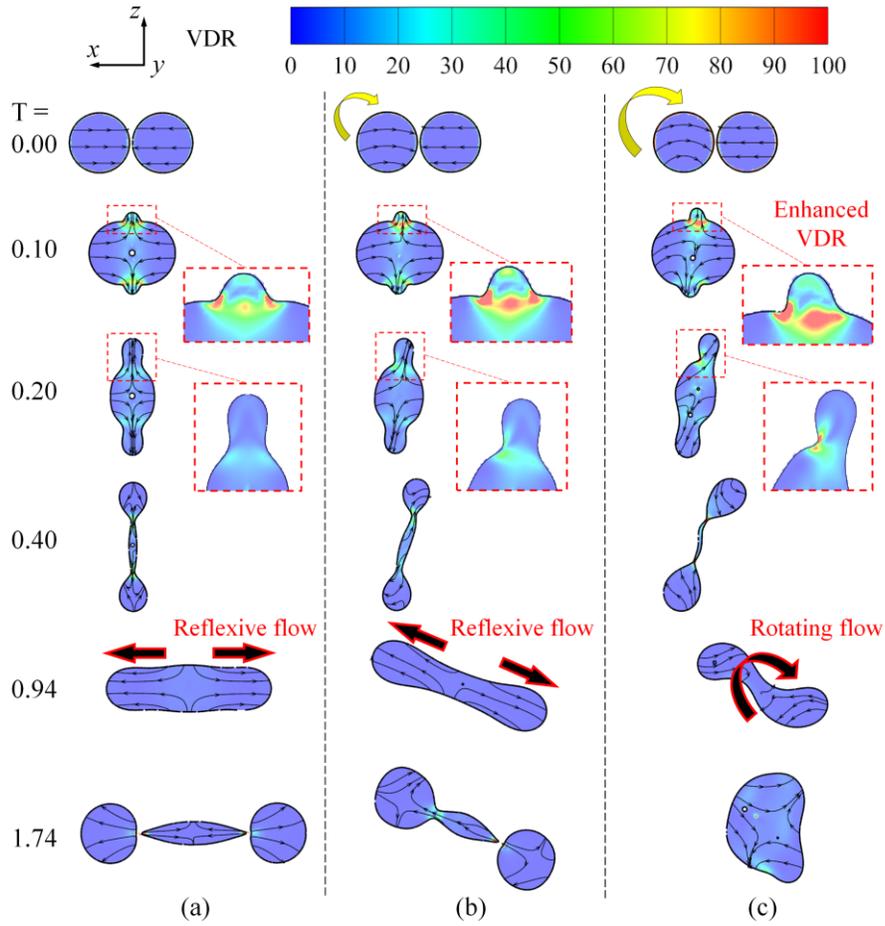

FIG. 6. Contour of the viscous dissipation rate (VDR) and streamlines for the three cases of reflexive separation at $B = 0.0$ that shown in Fig. 5(a-c) with (a) $\omega_0 = 0$, (b) $\omega_0 = 3$, and (c) $\omega_0 = 6$.

## B. Separation mechanisms affected by $\omega_0$

For the collisions at small $B$, the reflexive separation is suppressed by increasing $\omega_0$. This observation can be explained as the enhanced viscous dissipation[32] induced by the interaction between the spinning and translational motions of droplets. Specifically, the viscous dissipation rate (VDR) [30] is calculated by

$$\text{VDR} = \frac{\mu_l}{2}(\partial_j u_i + \partial_i u_j)^2 = 2\mu_l D_{ij}^2 \tag{8}$$

and shown in Fig. 6. It is clearly seen that the viscous dissipation is enhanced around the outwardly deformation interface, which also has large rotational velocity induced by the droplet spin. In addition, the tendence of reflexive flow accounting for the reflexive separation is reduced by the enhanced inner rotating flow. If the kinetic energy of the droplet reflexive motion is not sufficiently large to overcome the surface tension energy, it will eventually be dissipated during the droplet oscillation process, and consequently leading to droplet coalescence.

It has been shown in Fig. 5(d-f) that the stretching separation is promoted as increasing $\omega_0$ for the collisions at intermediate $B = 0.4$. This is because the total angular momentum ($L_t = L_o + L_{s1}$) of the merged droplet is increased so as to enhance droplet stretching and hence droplet separation. Specifically, the vorticity $\omega_y$ and streamlines in the *y*-direction are shown in Fig. 7, and they can reflect the local spinning motion during droplet deformation. It is seen that the vorticity distribution shows a point symmetry with respect to the origin for the non-spinning case, whereas the point symmetry is broken due to the nonzero initial vorticity of the spinning droplet. The liquid mass from the spinning droplet has a larger kinetic energy (KE) than that from the initially non-spinning droplet, as shown by the embedded contour of KE at T=0.22. The locally enhanced KE on one side of the merged droplet tends to elongate the merged droplet to form a longer ligament, as shown in Fig. 7(c). The enhanced local stretching flow owing to the droplet spin can overcome the surface tension to separate the ligament.

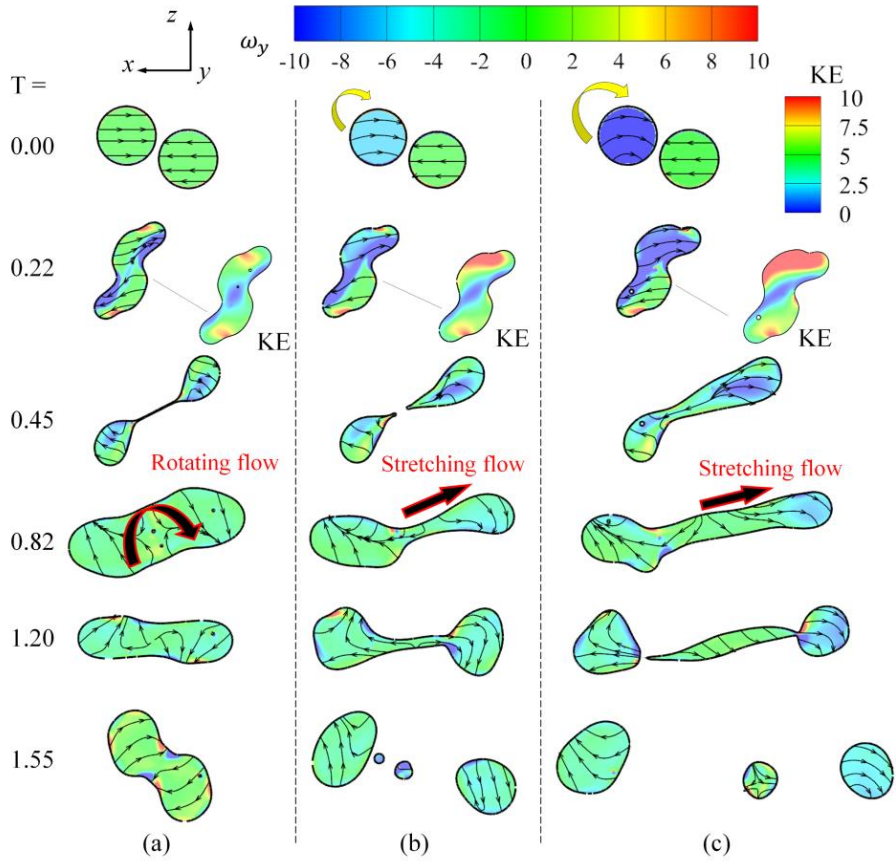

FIG. 7. Contour of the vorticity and kinetic energy (KE) fields and streamlines for the three cases of stretching separation at $B = 0.4$ that shown in Fig. 5(d-f) with (a) $\omega_0 = 0$, (b) $\omega_0 = 3$, and (c) $\omega_0 = 6$.

For the collisions at $B = 0.9$ that are shown in Fig. 5(g-i), the spinning motion can lead to a longer ligament and then breakup into more satellite droplets as increasing $\omega_0$. This can be explained by the same mechanism of the stretching separation at intermediate $B$ as that the stretching effect is enhanced by the droplet self-spin. It is interesting to see in Fig. 8 that the liquid mass from the initial spinning droplet can have a residual spin angular momentum. The vorticity of the residual spin is enhanced as increasing $\omega_0$.

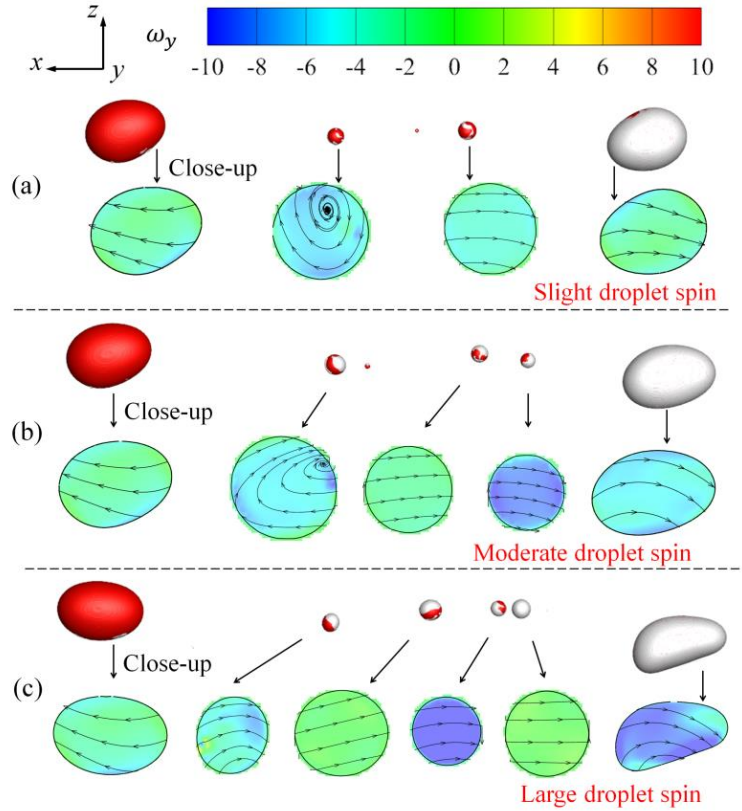

FIG. 8. Contour of the vorticity field and satellite droplet formation upon the three cases of nearly grazing collision at $B = 0.9$ that shown in Fig. 5(g-i) with (a) $\omega_0 = 0$, (b) $\omega_0 = 3$, and (c) $\omega_0 = 6$.

## IV. EFFECTS OF DROPLET SPIN AND ITS CHIRALITY

### A. Influences of droplet spin chirality on $We - B$ regime nomogram

It is noted that the spinning motion of the droplet is so set up that its angular momentum has the same direction with the orbital angular momentum. Consequently, their synergetic influence on the droplet collision is through the scalar total angular momentum ($L_o + L_{s1}$). It is a natural question to ask what is the influence of the droplet spinning direction, which apparently can change the total angular momentum $\boldsymbol{L}_t$ and hence the rotational energy of the collision system. Considering the angle ($0 \leq \psi \leq \pi$) between the spin and the orbital angular momentum, we can infer that the influence should be minimal at $\psi = \pi/2$ (orthogonality) and be maximal at either $\psi = 0$ (corresponding to the cases discussed so far) or $\psi = \pi$, which will be discussed as follows.

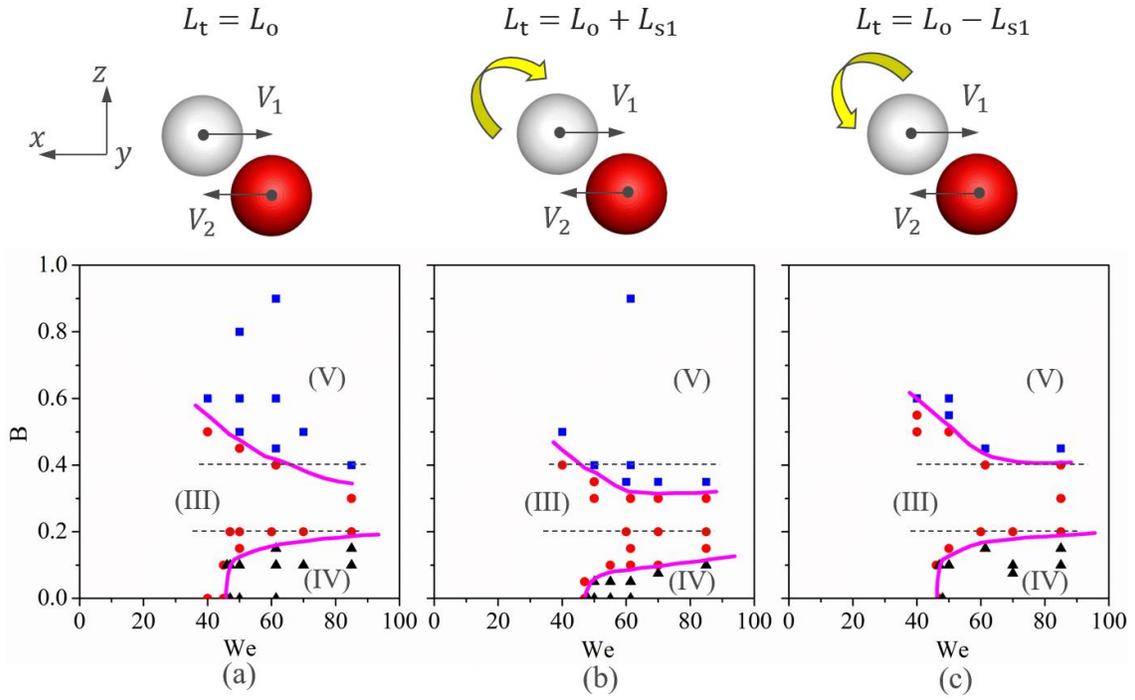

FIG. 9. Influence of chirality of droplet spin on the transition of coalescence-separation boundary at $Oh = 0.028$ for the case of (a) two non-spinning droplets and the cases of one spinning droplet with (b) $\boldsymbol{\omega_0}\hat{y} = -3$ and (c) $\boldsymbol{\omega_0}\hat{y} = 3$.

One of the most prominent effects of the spin chirality is its effect on the stretching separation, as shown in the regime nomogram in Fig. 9. Compared with the non-spinning cases, as shown in Fig. 9(a), the stretching separation-coalescence regime boundary moves down to smaller $B$ in Fig. 9(b) for the "positive" spin, which results in the addition of the orbital and spin angular momentums ($L_o + L_{s1}$), and the regime boundary moves up to larger $B$ in Fig 9(c) for the "negative" spin, which results in the counteraction of the angular momentums ($L_o - L_{s1}$). This finding is consistent with the above discussions for Fig. 7 that a larger value of $L_t$ could generate a more significant local stretching flow and therefore a longer liquid ligament leading to a stretching separation even if at a smaller $B$. Similarly, a smaller $L_t$ reduces the local stretching inertia, and a larger $B$ is therefore necessary for stretching separation of the merged droplet.

Another prominent effect of the spin chirality is that on the reflexive separation. Specifically, the reflexive separation is suppressed by the "positive" spin as seen in Fig. 9(b), whereas is negligibly influenced by the "negative" spin as seen in Fig. 9(c). This can be explained from a perspective of energy budget by using a representative case of the nearly head-on collision at $We = 61.4$ and $B = 0.15$, as shown in Fig. 10. As discussed in the preceding section, the reflexive separation is influenced by the competition between the viscous dissipation within the reflexive flow and the rotating flow that diminishes the reflexive flow. Although the total energy (TE) for three cases is approximately the same, the kinetic energy (KE) and surface energy (SE) for the case (b) with $L_t = L_o + L_{s1}$ are different from others because the reflexive separation does not happen for this case. It is also noted that there are minor differences of the total viscous dissipation rate, TVDR(T)=$\int_V$ VDR $H(c-1)\,dV$, and the time accumulated viscous dissipation energy, TVDE= $\int_0^T$ TVDR(T)$dT$ [30] for the three cases. It is inferred that the viscous dissipation is not a decisive factor accounting for the influence of the spin chirality on the reflexive separation.

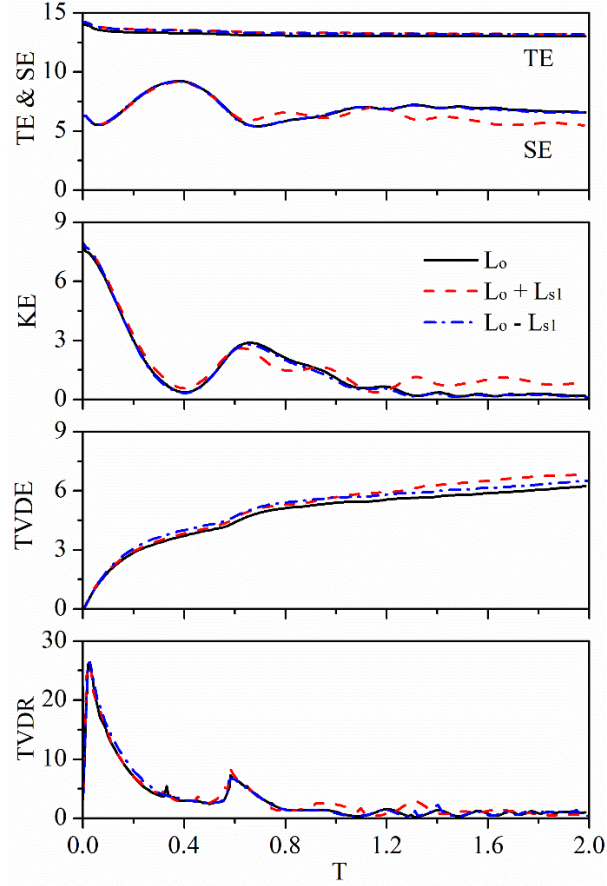

FIG. 10. Energy budget of the cases for the nearly head-on collision at $We = 61.4$ and $B = 0.15$ that are extracted from Fig. 9, in which the total energy (TE), the surface energy (SE), the kinetic energy (KE), the total viscous dissipation energy (TVDE), and the total viscous dissipation rate (TVDR) that are all nondimensional.

The vorticity field that reflects the local inner flow is illustrated in Fig. 11. Compared with the reflexive separation in case (a) and (c), the droplet spin in case (b) leads to a non-uniform vorticity distribution on the round head of the ligament. The vorticity diffusion due to the vorticity nonuniformity would induce a net rotating flow from the head of the ligament to it center and therefore strengthen the ligament from being separated. For the case (c), the slightly non-uniform vorticity distribution is insufficient to induce a prominent rotating flow to counteract the reflexive flow so that the reflexive separation occurs.

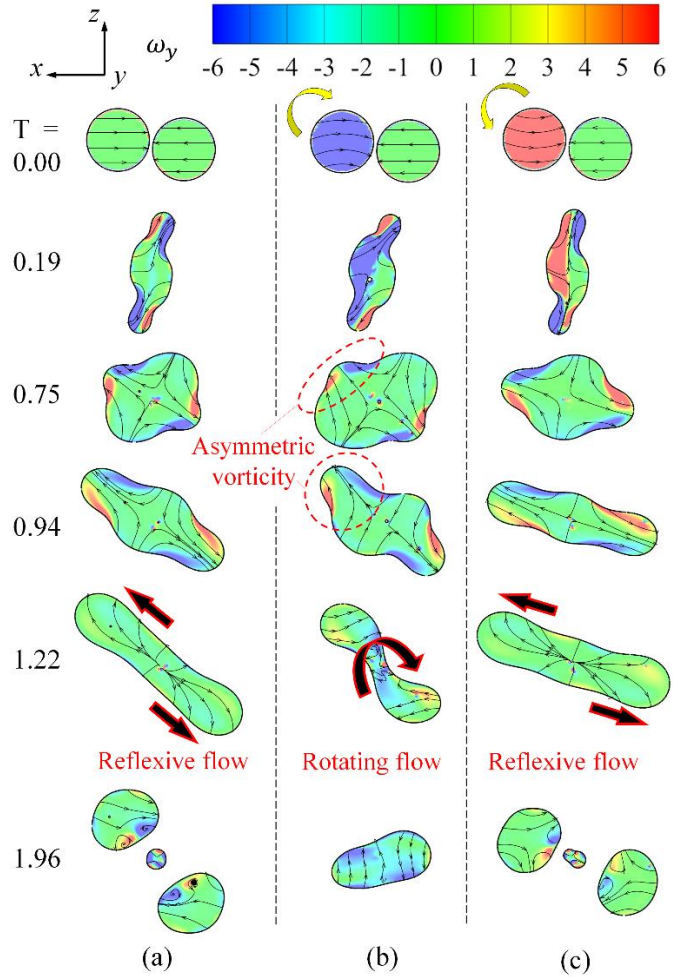

FIG. 11. Contour of the vorticity filed and streamlines for droplet reflexive separation that shown in Fig. 10 for the case of (a) two non-spinning droplets and the cases of one spinning droplet with (b) $\boldsymbol{\omega_0}\hat{y} = -3$ and (c) $\boldsymbol{\omega_0}\hat{y} = 3$.

## B. Separation of two spinning droplets with different chirality

For a complete discussion about the chirality effects, the collisions between two spinning droplets with either opposite or same spinning directions are examined, based on two representative cases of $B = 0.0$ and $B = 0.4$ from Fig. 5 (at $We = 61.4$, $Oh = 0.028$, and $\omega_0 = 3$).

For the head-on collision between two spinning droplets with opposite chirality in Fig. 12(a), the droplet deformation and vorticity contour are of a "mirror symmetry" with respect to the interaction (y-z) plane because of $L_t = 0$. The reflexive separation is observed, although appearing a shorter ligament and a smaller satellite droplet when compared with the case shown in Fig. 4(d) for a spinning droplet and a non-spinning droplet. For collision droplets with same chirality in Fig. 12(b), the reflexive separation is suppressed because both the rotating flow and rotational kinetic energy are enhanced by the increase of angular momentum $L_t = 2L_{s1}$. It consolidates our understanding that the rotational kinetic energy associated with the angular momentum can suppress the reflexive separation for the nearly head-on collision.

For the off-center collision between two spinning droplets shown in Fig. 13, the situation is more complex than that of the head-on collision in Fig. 12 due to the nonzero $L_o$. Thus, we compare two different situations with the reference spin of droplet $O_1$ being clockwise ("positive") or anticlockwise ("negative") that can increase or decrease total angular momentum, respectively. In addition, to characterize the roles of initial local stretching flow on separation for each droplet, the value of orbital angular momentum $L_o$ has been manually divided into two identical components of $L_{o1}$ and $L_{o2}$ for each droplet to facilitate the discussion.

By comparing the opposite droplet spin direction in Fig. 13(a) and the collision between two non-spinning droplets in Fig. 5(d), they have the same $L_t = L_o$ however the stretching separation is observed in Fig. 13(a). This is because the initial total angular momentum for droplet $O_1$ is $L_{o1} + L_{s1}$, which leads to an enhanced local stretching flow and promotes a long ligament formation and thereby stretching separation. Conversely, $L_{o2} - L_{s2}$ for droplet $O_2$ causes a reduced local stretching flow that suppresses the stretching separation, and consequently the ligament pinched-off occurs only on one side and no satellite droplet is formed. For the case shown in Fig. 13(b), the local stretching flow of two spinning droplets are both enhanced so as the stretching separation is readily to occur with a longer ligament, while for the case shown in Fig. 13(c), the local stretching flow of two droplets are both reduced so that no separation occurs. All these observations consolidate the finding that the local stretching flow associated with the angular momentum play important roles in promoting or suppressing the stretching separation for off-center collisions.

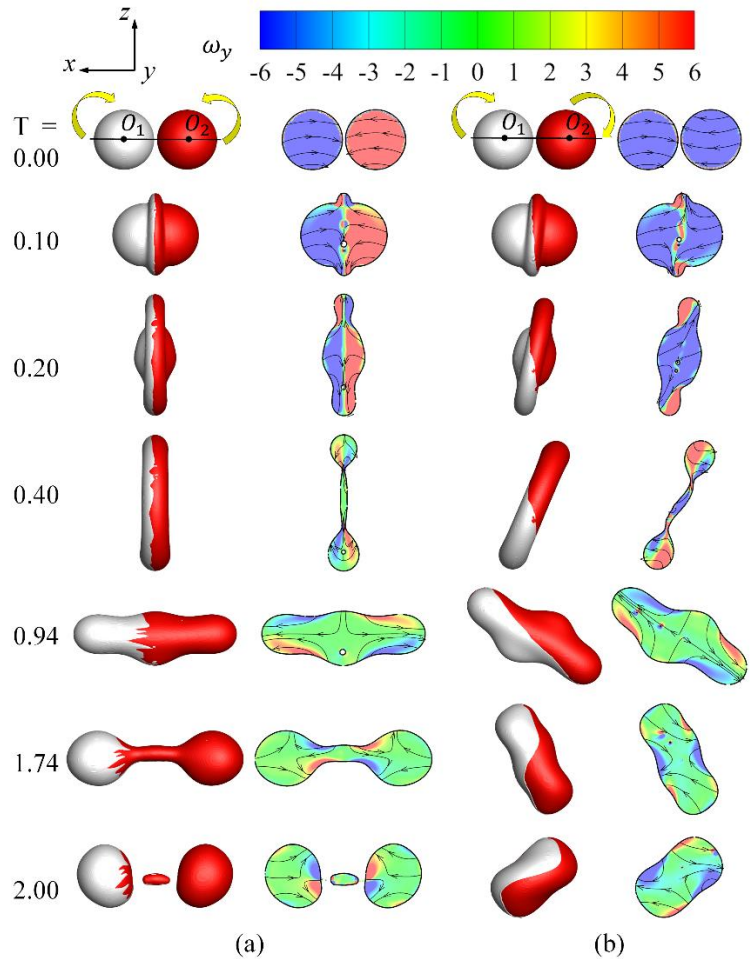

FIG. 12. Droplet deformation and vorticity contour for a head-on collision between two spinning droplets at $We = 61.4$, $Oh = 0.028$ and $\omega_0 = 3$. The spin motions have the (a) opposite direction and (b) the same direction.

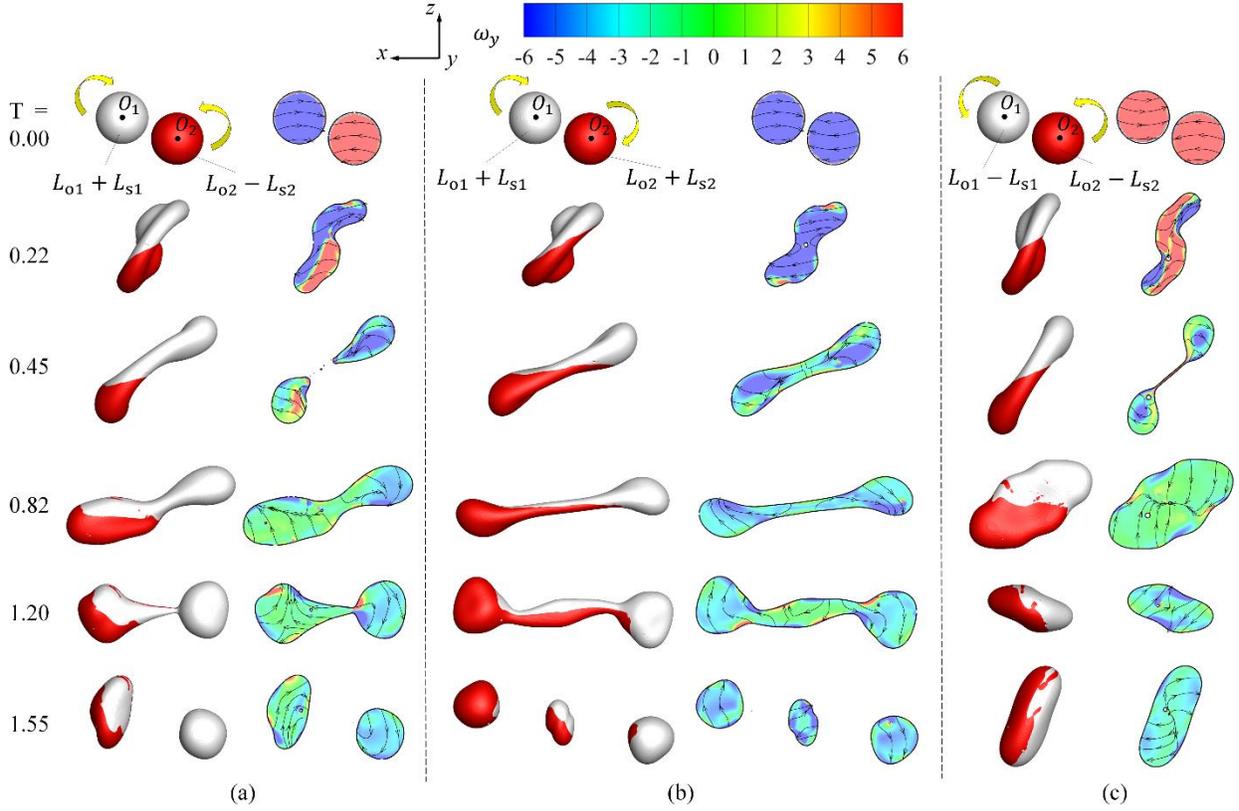

FIG. 13. Droplet deformation and vorticity contour for an off-center collision between two spinning droplets at $We = 61.4$, $Oh = 2.8 \times 10^{-2}$, $B = 0.4$ and $\omega_0 = 3$. The spin motions have (a) opposite direction (droplet spin of $O_1$ is clockwise), (b) the same direction but droplet spin of $O_1$ is clockwise, and (c) the same direction but droplet spin of $O_1$ is anticlockwise.

## V. THEORETICAL MODEL FOR SPIN CHIRALITY EFFECTS ON DROPLET SEPARATION

A theoretical model for the criterion of reflexive and stretching separation based on the energy balance analysis for the collision between two non-spinning droplets is given in the Appendix. Here, we further propose an extended version of the model by accounting for the droplet spin chirality effects. It should be emphasized that the proposed model is used to qualitatively interpret the spinning effects on separation. A quantitatively predictive model requires comprehensive parametric study on all controlling parameters and indeed merits future study.

Apart from the inner reflexive and stretching flows induced by the off-center collision, the droplet spin motion induces an internal rotating flow with the velocity $U_{\text{ro}} \sim R\omega_0$, which results in an additional viscous dissipation in Stage II proportional to $\mu(U_{\text{ro}}/R)^2 \left(\frac{4}{3}\pi R^3\right)\sqrt{\rho_l R^3/\sigma}/8\pi R^2 \sigma$, and its non-dimensional form is given by

$$\Phi_{\text{II,ro}} = \beta_5 Oh W e_s \tag{9}$$

where $We_s = \rho_l D_l (R\omega_0)^2/\sigma$ is the defined spin Weber number. The total viscous dissipation during the stage II is given by $\Phi_{\text{II}} = \Phi_{\text{II,r}} + \Phi_{\text{II,s}} + \Phi_{\text{II,ro}}$, which degenerates to (A3) for the non-spinning cases. Regarding the energy balance analysis, the droplet spin motion effects the additional rotational kinetic energy (RKE$_0$) given by

$$\text{RKE}_0 = \frac{\boldsymbol{L_{s1}}^2}{2I_{y1}} = \frac{We_s}{60} \tag{10}$$

where $\boldsymbol{L_{s1}}$ is the spin angular momentum given by Eq. (7a).

The droplet spin motion influences the local stretching flow. As we have seen in Fig. 7, due to the asymmetric kinetic energy distribution during the droplet rotation and separation stage, the ligament length can be either enhanced or reduced on the side with the mass liquid connected to the spinning droplet by increasing or decreasing the total angular momentum, respectively. Thus, the local stretching flow $U_{\text{s,eff}}$ influenced by the droplet spin motion requires to be modified as $U_{\text{s,eff}} \sim U_s + \beta_6 \eta R\omega_0$, where $\eta = (\boldsymbol{L_o} \cdot \boldsymbol{L_{s1}})/|\boldsymbol{L_o} \cdot \boldsymbol{L_{s1}}| = 1$ (or $-1$) corresponding to that $\boldsymbol{L_o}$ and $\boldsymbol{L_{s1}}$ have the same (or opposite) direction.

Following to the derivations expatiated in the Appendix, we can obtain the modified moment of inertia and effective rotational kinetic energy (E$_{\text{r,eff}}$) of the merged droplet by replacing the stretching velocity $U_s$ with the effective local stretching velocity $U_{s,eff}$. Consequently, Eq. (A7) can be modified as

$$(1-\alpha)We + \alpha WeB^2$$
$$> \beta_1 Oh\left(1 + \beta_2/\beta_1\sqrt{WeB^2} + \beta_5/\beta_1 We_s\right) + \gamma \quad (11)$$
$$+ \frac{\beta_4 We_{eff}B^2}{(\beta_3 We_{eff}B^2)^{-2} + (\beta_3 We_{eff}B^2)^2} - \frac{We_s}{60}$$

where $We_{eff}B^2 = \left(\sqrt{WeB^2} + \beta_6\eta\sqrt{We_s}\right)^2$ is the effective tangential component of stretching inertia. By using the same fitting coefficients given in the Appendix ($\alpha = 0.5$, $\beta_1 = 30$, $\gamma = 15$, $\beta_2 = 30$, $\beta_3 = 0.3$, and $\beta_4 = 20$) for the collisions between two non-spinning droplets, Fig. 14 compares the predicted boundaries with different values of $\beta_5$ and $\beta_6$. It is seen that the predicted boundaries are not sensitive to the variation of $\beta_5$ and $\beta_6$ in wide ranges, and $\beta_5 = 3$ and $\beta_6 = 0.1$ produce the best fitting to the present computational results. As shown in Fig. 14(a) and 14(c), by increasing $\beta_5$, the boundary shifts towards a larger $We(1 - B^2)$ owing to the increased viscous dissipation induced by the droplet spin motion. By increasing $\beta_6$, the boundary shifts towards a smaller $WeB^2$ as shown in Fig. 14(b) because the reflexive separation is suppressed while the stretching separation is promoted by the "positive" spin (with $\eta = 1$) as seen in Fig. 9(b); and conversely the boundary shifts towards a larger $WeB^2$ as shown in Fig. 14(d) because of the "negative" spin (with $\eta = -1$) as seen in Fig. 9(c).

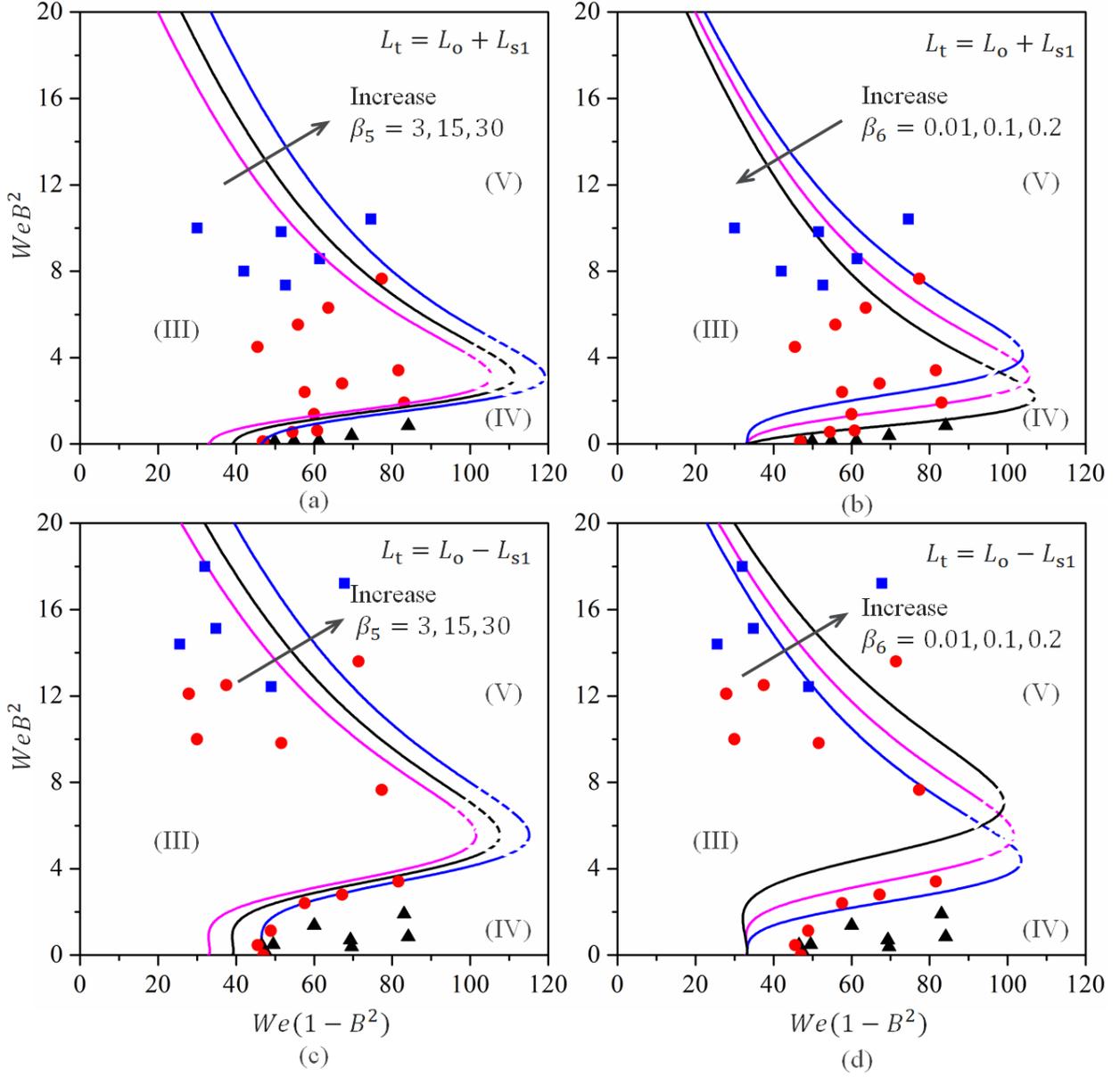

FIG. 14. Comparison of the predicted boundary for separations upon the collision between a spinning and a non-spinning droplet in Fig. 9. For sensitivity analysis, different values of $\beta_5$ and $\beta_6$ in large ranges are shown as (a) $\beta_5$ and (b) $\beta_6$ for $L_t = L_o + L_{s1}$; (c) $\beta_5$ and (d) $\beta_6$ for $L_t = L_o - L_{s1}$.

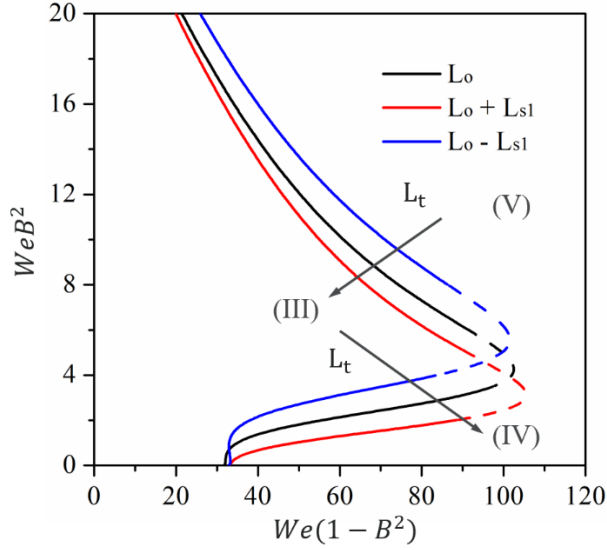

FIG. 15. Comparison of the predicted boundaries of three representative cases in Fig. 9, (a) two non-spinning droplets, (b) $\boldsymbol{\omega_0}\hat{y} = -3$, and (c) $\boldsymbol{\omega_0}\hat{y} = 3$.

Figure 15 further compares the predicted boundaries for three representative cases shown in Fig. 9. As increasing $L_\mathrm{t}$, the lower branch of a boundary shifts towards a larger $We(1-B^2)$, while the upper branch shifts towards a smaller $WeB^2$. This indicates that the present theoretical model can qualitatively predict the chirality of spinning effects on the coalescence-separation transition that increasing the total angular momentum suppresses the reflexive separation but promotes the stretching separation.

## VI. CONCLUSIONS

A computational study on droplet separation that is influenced by the droplet spin motion was investigated based on a validated Volume-of-Fluid method. The results show that droplet spin motion influences the reflexive and stretching separation in different ways. For the nearly head-on collisions, as increasing the spin angular momentum of the spinning droplet, the reflexive separation tends to be suppressed owing to the enhanced viscous dissipation and the induced inner rotating flow that diminishes the reflexive flow. For the collisions at larger impact parameters, the stretching separation is promoted by the local enhanced stretching flow. For the nearly grazing collisions, a longer ligament is formed and breaks into more satellite droplets, in which the liquid mass from the initial spinning droplet can reserve partial spin angular momentum.

The key physics underlying the rich phenomena is that the interplay between the initial spin angular momentum of spinning droplet and the orbital angular momentum. This can be further illustrated by changing the chirality of droplet spin, which results in an augmented or reduced total angular momentum. For the nearly head-on collisions, it is found that the inner rotating flow is a decisive factor accounting for the influence of the spin chirality on the reflexive separation. For the collisions at larger impact parameters, the stretching separation depends on the local stretching flow, which is enhanced or reduced by the increase or decrease of total angular momentum.

A theoretical model for the reflexive and stretching separation from the aspect of energy balance analysis is proposed for the collisions involving spinning droplets. It can qualitatively reflect the non-monotonicity of the separation varying with the impact parameter from reflexive separation to coalescence and to stretching separation. The key component of the model responsible for the non-monotonicity is the rotational kinetic energy, which plays an important role in affecting the droplet separation but was not considered in previous modelling studies. By taking into account of the contribution of droplet spin in the rotational kinetic energy, the model can also qualitatively explain the effects of droplet spin chirality on suppressing the reflexive separation while promoting the stretching separation by increasing the total angular momentum.

The off-center collision between spinning droplets with arbitrary polar and azimuthal angles of their spin axes is an apparent and necessary extension of the present work. But such as future work is certainly more complex than the present one by breaking more symmetries that help simplify the present problem. Besides, the experimental confirmation of the present computational results is of significance but challenging, because it may require some innovations of the current experimental techniques in generating and visualizing spinning droplets.

**ACKNOWLEDGMENTS**

This work was supported by the China Postdoctoral Science Foundation (2020M680690), by Hong Kong RGC/GRF (through Grant No. PolyU 152188/20E) and by the Hong Kong Polytechnic University DGRF (through Grants No. G-UAHP).

**APPENDIX: A THEORETICAL MODEL FOR UNIFIED SEPARATION OF NON-SPINNING DROPLETS**

In previous studies on non-spinning droplet collision[4, 5], the energy balance analysis were applied to model reflexive separation and stretching separation of two non-spinning droplets, respectively. In this appendix, a unified model will be proposed for droplet separation of two non-spinning droplets at arbitrary $B$. As shown in Fig. A1, the entire collision process is divided into three distinct stages: (I) the impacting stage from the instant upon droplet collision to that when the normal component of impact inertia (associated with the normal component of impact velocity $U_n = U\sqrt{1-B^2}$ ) has been completely converted into surface energy and viscous dissipation; (II) the rebounding stage (owing to the surface energy converted into reflexive inertia) for reflexive separation and the stretching stage (continuing stretching owing to the tangential component of impacting inertia that is associated with the tangential component of impact velocity $U_s = UB$) for stretching separation, respectively; and (III) the rotation and separation stage for the rotating droplet (owing to the nonzero angular momentum of the collision system), which appears different outcomes with varying $B$.

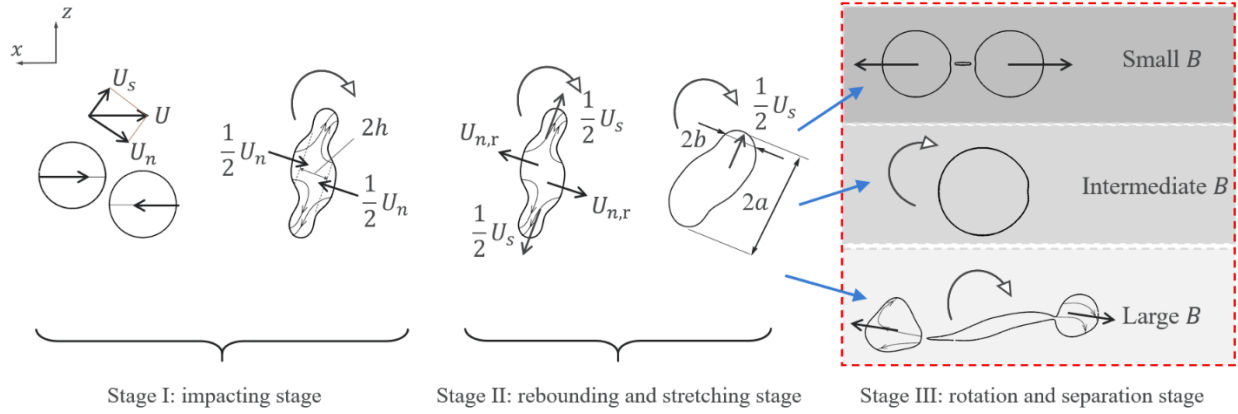

FIG. A1. Schematic of the model adopted in analyzing droplet reflexive and stretching separation for the collision between two non-spinning droplets.

In stage I, the characteristic time can be estimated to be the order of $2R/U_n$; the order of magnitude of the strain rate in the stagnation flow region is $U_n/h$, where the characteristic width $h \sim 2\,\mu_l/\rho_l U_n$ is obtained by using the stress balance relation $\frac{1}{2}\rho_l U_n^2 \sim \mu_l\, U_n/h$; the volume of the stagnation flow region is of the order of $(2h)(\pi R^2)$. Thus, most of the viscous dissipation in stage I is due to the viscous dissipation in the stagnation flow region, proportional to $[\mu_l(U_n/h)^2(2\pi h R^2)(2R/U_n)]/8\pi R^2 \sigma$, and its non-dimensional form is given by,

$$\Phi_\mathrm{I} = \alpha We(1 - B^2) \tag{A1}$$

where $0 < \alpha < 1$ is a proportionality factor to be determined by fitting experiments. Jiang *et al.*[4] found that $\alpha = 0.5$ is a universal coefficient for the cases of $B = 0$.

In stage II of the rebounding or stretching stage, the viscous dissipation is caused by both the reflexive flow and the stretching flow. The reflexive flow is driven by the surface tension and satisfies the relation of $\frac{1}{2}\rho_l U_{n,r}^2 \sim \sigma/R$, and the stretching flow is approximately estimated by $U_s = UB$. The characteristic time for reflexive separation is estimated to be the order of the natural oscillation time $t = \sqrt{\rho_l R^3/\sigma}$, whereas that for stretching separation being the order of $2R/U_s$. Thereby, the viscous dissipation due to the reflexive flow is proportional to $\mu_l(U_{n,r}/R)^2 \left(\frac{4}{3}\pi R^3\right)\sqrt{\rho_l R^3/\sigma}/8\pi R^2 \sigma$, and its non-dimensional form is given by

$$\Phi_{\text{II,r}} = \beta_1 Oh \tag{A2a}$$

and the viscous dissipation due to the stretching flow is proportional to $\mu_l(U_s/R)^2 \left(\frac{4}{3}\pi R^3\right)(2R/U_s)/8\pi R^2 \sigma$, and its non-dimensional form is given by

$$\Phi_{\text{II,s}} = \beta_2 Oh\sqrt{WeB^2} \tag{A2b}.$$

The total viscous dissipation during the stage II is approximated by

$$\Phi_{\text{II}} = \Phi_{\text{II,r}} + \Phi_{\text{II,s}} \tag{A3}.$$

Qian and Law[5] suggested $\beta_1 = 30$ by fitting their experimental data for head-on collisions.

In stage III of the rotating and separation stage, the off-center collision-induced rotation motion would suppress the droplet separation, owing to the conversion of initial orbital angular momentum $\boldsymbol{L}_o$ into the spin angular momentum $\boldsymbol{L}_s$ of the merged droplet. As shown in Fig. A1, the critical droplet deformation at small $B$ is approximately two spheres connected by a short ligament; that at intermediate $B$ is approximately a larger sphere; and that at large $B$ is approximately two spheres connected by a long ligament. The characteristic length of the round head of stretching ligament is $2b$, as shown in Fig. A1, which can be estimated by the balance between the surface tension pressure and the stagnation pressure as $\sigma/b \sim \frac{1}{2}\rho_l U_s^2$, yielding $b \sim 2\sigma/\rho_l U_s^2$ and $a \sim R^2/b$ by the volume conservation.

The moment of inertia of the merged droplet (or approximate a stretching ligament) based on the spinning axis (y-axis) is estimated by

$$I_y = \frac{1}{5}m(a^2+b^2) \sim \rho_l R^5 \left[\frac{1}{(\beta_3 WeB^2)^2} + (\beta_3 WeB^2)^2\right] \quad (A4)$$

The proportionality factor $\beta_3$ appears symmetrically in (A4) for the mathematical convenience. Physically, $I_y$ varies non-monotonically with $WeB^2$ and reaches its minimum value at $WeB^2 \sim 1/\beta_3$. For non-spinning droplet collision, $\mathbf{L}_t = \mathbf{L}_o$ is given in Eq. (7). Consequently, the rotational kinetic energy ($E_r = L_t^2/2I_y$) of the merged droplet, normalized by $8\pi R^2 \sigma$, can be given by

$$E_r = \frac{\beta_4 WeB^2}{(\beta_3 WeB^2)^{-2} + (\beta_3 WeB^2)^2} \quad (A5)$$

The physical criterion from the perspective of energy balance is that the separation occurs if the viscous dissipation (in Stage I and II) and $E_r$ in Stage III are smaller than the excessive kinetic energy so that the merged droplet cannot hold. This criterion can be expressed in a non-dimensional form as

$$E_{k0} + E_{s0} > \Phi_I + \Phi_{II} + E_r + E_s \quad (A6)$$

in which $E_{k0}$ is the non-dimensional initial impacting kinetic, $E_{s0}$ the initial surface energy, and $E_s$ the surface energy of the merged droplet at the instant that about to separate. Following Qian and Law[5], we defined the surface energy addition $\gamma = E_s - E_{s0}$, where $\gamma = 15$ for head-on collisions.

Consequently, by substituting Eq. (A1), (A3) and (A5) into Eq. (A6), we obtain the transition boundary for separation as

$$(1-\alpha)We + \alpha WeB^2$$

$$> \beta_1 Oh\left(1 + \beta_2/\beta_1\sqrt{WeB^2}\right) + \gamma \tag{A7}$$

$$+ \frac{\beta_4 WeB^2}{(\beta_3 WeB^2)^{-2} + (\beta_3 WeB^2)^2}$$

For head-on collisions ($B = 0$), Eq. (A7) can be degenerated into

$$(1-\alpha)We_c > \beta_1 Oh + \gamma \tag{A8}$$

which is exactly the same as the equation derived by Qian and Law[5].

To determine proportionality factors, we applied the previous results and scaling analysis. Specifically, $\alpha = 0.5$ is consistent with previous results in[4, 5]; $\beta_1 = 30$ and $\gamma = 15$ were suggested by Qian and Law[5] by fitting experiment for the head-on collision; $\beta_3 = 0.3$ is consistent with the estimate of $\beta_3 \sim (WeB^2)^{-1} \sim O(10^{-1})$ for droplet collisions in the parameter range of $We \sim O(10^2)$ and $B^2 \sim O(10^{-1})$; and $\beta_4 = 20$ accords with that $\beta_4 \sim O(10)$, which follows from the relation $\beta_4 WeB^2 \sim We$ because the initial impact kinetic energy ($E_{k0}$) and $E_r$ have the same order of magnitude.

The predicted boundary for reflexive and stretching separations by using (A7) with the above factors is shown in Fig. A2. It is seen that (A7) can qualitatively reflect the trends of reflexive separation and stretching separation with varying $B$ and $We$. It is noted that a part of the predicted boundary around a turning point is shown by a dashed curve because droplet coalescence actually occurs. It is a common phenomenon in many physics problems[45, 46] that a turning point separates the boundary to two stable branches (here, the lower branch is for reflexive separation and the upper branch for stretching separation), and that unstable nature of the problem around the turning point often requires additional physics. The present model cannot predict the occurrence of droplet coalescence because it does not account for a few key physics, such as gas film drainage, rarified gas effects, and van der Waals force. A unified model that is able to predict both separation and coalescence certainly merits future studies.

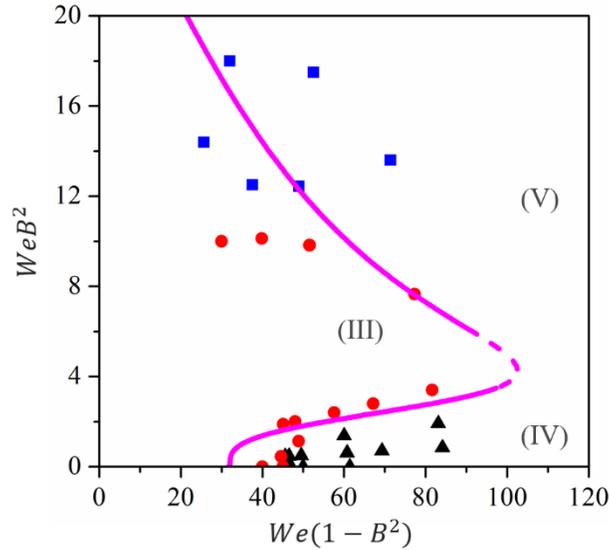

FIG. A2. Predicted boundary for the reflexive and stretching separation of two non-spinning droplets in the parameter space $We(1-B^2) \sim WeB^2$. The dashed curve indicates the occurrence of droplet coalescence, which cannot be predicted by the present model.

To examine the sensitivity of the prediction by (A7) to the proportionality factors, particularly $\beta_2$, $\beta_3$, and $\beta_4$, we further compare the predicted boundaries by using different values of $\beta_2$, $\beta_3$, and $\beta_4$, as shown in Fig. A3. All the boundary curves show the same qualitative trend by varying the proportionality factors in a wide range, and these results validate the estimated orders of magnitude of these proportionality factors in the present model. Specifically, the boundary curves are slightly shifted towards a larger $We(1-B^2)$ by increasing $\beta_2$ owing to the increased viscous dissipation. Decreasing $\beta_3$ or increasing $\beta_4$ tend to enhance the amount of rotational kinetic energy so as droplet separation is suppressed with enlarged regime of droplet coalescence (III). This indicates that the rotational kinetic energy plays an important role in affecting the droplet separation.

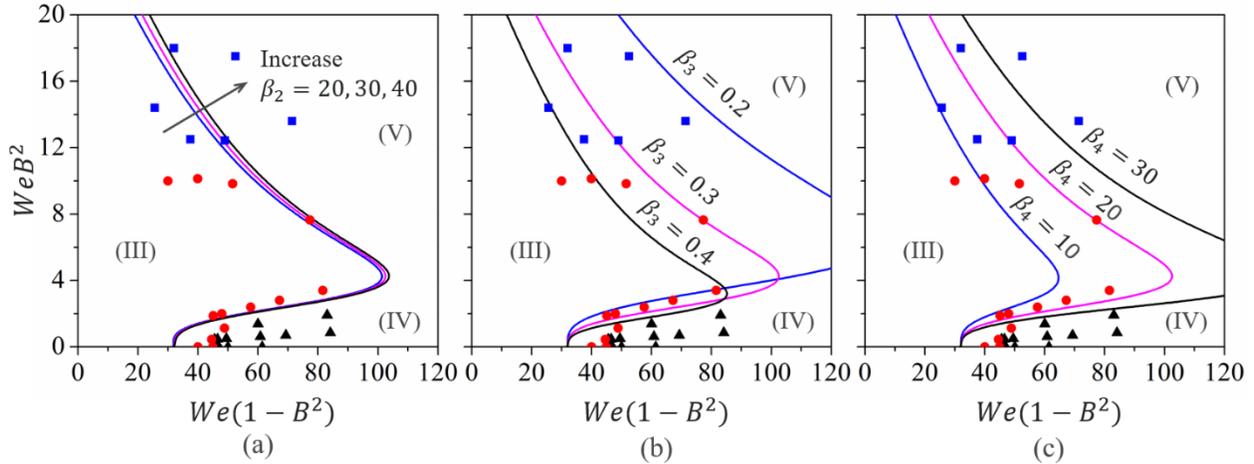

FIG. A3. Comparison of the predicted boundaries with different proportionality factors: (a) $\beta_2$, (b) $\beta_3$, and (c) $\beta_4$.